\documentstyle[12pt,axodraw]{article}
\newcommand{\be}{\begin{equation}}
\newcommand{\ee}{\end{equation}}
\newcommand{\bea}{\begin{eqnarray}}
\newcommand{\eea}{\end{eqnarray}}

\newcommand{\gm}{\gamma}

\newcommand{\dl}{\delta}

\newcommand{\ep}{\epsilon}

\newcommand{\dd}{\mbox{d}}

\newcommand{\nn}{\nonumber}

\begin{document}
\parindent=1.5pc

\begin{titlepage}
\rightline{NPI MSU 2000--4/608}
\rightline{hep-ph/0001192}
\rightline{January 2000}
\bigskip
\begin{center}
{{\bf
Equivalence of Recurrence Relations for \\
Feynman Integrals with the Same Total Number of \\
External and Loop Momenta} \\
\vglue 5pt
\vglue 1.0cm
{ {\large P.A. Baikov\footnote{E-mail: baikov@theory.npi.msu.su}
and
V.A. Smirnov\footnote{E-mail: smirnov@theory.npi.msu.su} } }\\
\baselineskip=14pt
\vspace{2mm}
{\em Nuclear Physics Institute of Moscow State University}\\
{\em Moscow 119899, Russia}
\vglue 0.8cm
{Abstract}}
\end{center}
\vglue 0.3cm
{\rightskip=3pc
 \leftskip=3pc
\noindent
We show that the problem of  solving  recurrence relations for $L$-loop
$(R+1)$-point  Feynman integrals within the method of integration by parts
is equivalent to the corresponding problem for  $(L+R)$-loop vacuum or
$(L+R-1)$-loop propagator-type integrals.
Using this property we solve recurrence relations for two-loop massless vertex
diagrams, with arbitrary numerators and integer powers of propagators in 
the case when two legs are on the light cone,  by reducing the problem to
the well-known solution of the corresponding recurrence relations for massless
three-loop propagator diagrams with specific boundary conditions.
\vglue 0.8cm}
\end{titlepage}

\section{Introduction}

The method of integration by parts (IBP) \cite{IBP,IBP2}
within dimensional regularization \cite{dimreg}
is based on integrating by parts in
dimensionally regularized Feynman integrals in loop momenta and
putting to zero surface terms. Although this procedure has been justified
only for off-shell Feynman integrals \cite{ChS}, no examples are known
where it breaks down so that it is successfully applied to Feynman
integrals on the mass shell, at threshold and to various
integrals in the loop momenta that arise in asymptotic expansions
of `standard' Feynman integrals in limits of momenta and masses.

The idea of the method is to use equations following from
IBP in order to derive recurrence relations that
reduce any Feynman integral from a given family to
a set of master integrals, e.g. with lowest powers of propagators.
If this program is fulfilled then the evaluation
of the given family of integrals reduces to algebraical manipulations.

In this  paper, we show that different problems of  solving  IBP relations
are in fact equivalent so that one can use known solutions to
provide solutions for other families of Feynman integrals.
We show that  the problem of  solving  IBP relations
for $L$-loop $(R+1)$-point Feynman integrals
is equivalent to the corresponding problem for  $(L+R)$-loop vacuum or
$(L+R-1)$-loop propagator-type integrals.

We apply  this property to reduce solution of the IBP relations for
two-loop massless vertex diagrams,
with arbitrary numerators and integer powers of propagators in the case
when two legs are on the light cone, to the
well-known solution \cite{IBP} for massless three-loop propagator-type
diagrams, but with specific boundary conditions.

\section{Equivalence of recurrence procedures}

Our goal is to prove the equivalence (in some special sense) of the
IBP relations for multi-loop Feynman  integrals with the same number of
independent kinematical invariants.
To do this  let us present (following \cite{NIM}, with updated
notation) the IBP relations in a modified form.
We start with $L$-loop vacuum Feynman integrals  with the maximum
possible number of the propagators, $N=L(L+1)/2$:
\begin{eqnarray}
 F(n_1,\ldots,n_N;d;m^2) =
\int \ldots \int \frac{\dd^dp_1\ldots \dd^dp_L}
{D_1^{n_1}\ldots D_N^{n_N}}
\equiv
(m^2)^{Ld/2 - \Sigma n_i} f(n_1,\ldots,n_N;d)
\, ,
\label{eqbn}
\end{eqnarray}
where $p_i$ (${i=1,\ldots,L}$) are loop momenta
and $D_a=A^{ij}_a p_i\cdot p_j -\mu_a m^2$ ($a=1,\ldots,N$;
summation over repeated indices is understood).
We use dimensional regularization \cite{dimreg} with $d=4-2\ep$.
The matrix $A^{ij}_a$ is supposed to be symmetrical with respect
to upper indices.
This maximal number $N$ of the propagators provides the possibility to
express any
scalar product of the loop momenta as a linear combination of the factors
in the denominator. Feynman integrals for specific graphs are
obtained for appropriate choices of the $N\times N$ matrices
$A^{ij}_a$.
Diagrams of practical interest which usually have less number of the
propagators can be considered as special cases with non-positive
powers of some propagators. If the number of the propagators
is greater than $N$, partial fractioning can be used to deal with integrals
with at most  $N$ propagators.

IBP relations are obtained \cite{IBP,IBP2}
by acting by the operator $(\partial/\partial p_i)\cdot p_k$ on
the integrand, putting the integral of the derivative to zero
and expressing all the
terms resulted from the differentiation through the initial family of
the integrals. Starting from (\ref{eqbn}) we obtain
\begin{eqnarray}
d\delta_{ik}  f(n_1,\ldots,n_N;d) &=&
2 \tilde{A}^{kl}_a ({\bf  I}^-_a+\mu_a)
A_b^{il}{\bf  I}_b^+  f(n_1,\ldots,n_N;d) \,,
\label{rr}
\end{eqnarray}
where
${\bf  I}^-_a f(\ldots, n_a,\ldots )\equiv f(\ldots, n_a-1,\ldots )$
and
${\bf  I}^+_a f(\ldots, n_a,\ldots )\equiv n_a f(\ldots, n_a+1,\ldots)$
(no summation in $n_a$ in the last relation ).
The elements $\tilde{A}^{kl}_a$ of the matrix inverse with respect to
 $A^{kl}_a$, i.e. with
$\tilde{A}^{ij}_a A^{kl}_a=(\dl_{ik} \dl_{jl}+\dl_{il} \dl_{jk})/2 $,
arise when expressing the
scalar products $p_i\cdot p_j$ in the numerator, which appear as
a result of the differentiation, in terms of the denominators $D_a$:
\be
p_k\cdot p_l = \tilde{A}^{kl}_a(D_a+\mu_a)
\,. \nn
\ee

Suppose now that the integrals depend in addition on $R$
external momenta $p_i$ ($L < i \leq L+R$)
and we are interested in their values at a particular point
$p_i\cdot p_k=\mu_{ik}m^2$. Without loss of generality we may assume
that $\det(\mu_{ik}) \neq 0$. Indeed, if this determinant is equal to zero
then the momenta $p_i$ at this point are linear dependent.
Let the rank of the matrix $\mu_{ik}$ be $R^\prime<R$. This means
that we can represent all expressions as functions of only $R^\prime$
external momenta so that we come back to the (slightly
simplified) original problem.

The number of independent propagators
(i.e. factors in the denominator)  which can be constructed from
$L$ loop and $R$ (independent) external momenta is $N_1=L(L+1)/2+LR$,
and the number of additional (`external') invariants is $N_2=R(R+1)/2$.
Let us expand the integrals in formal series (i.e. expand the integrands
in the corresponding Taylor series) in the external kinematical
invariants $ p_k\cdot p_l, \; k,l=L+1,\ldots,L+R$, or, in other words,
in the `denominator-like' objects $D_a,\,a=N_1+1,..,N_1+N_2$,
depending only on the external momenta:
\bea
 F(n_1,\ldots,n_N;d;m^2;p_{L+1},\ldots,p_{L+R )} &=&
\int \ldots \int \frac{\dd^dp_1\ldots \dd^dp_L}
{D_1^{n_1}\ldots D_N^{n_N}}
\nn \\&& \hspace*{-65mm}
= \sum_{n_{N_1},\ldots,n_{N_1+N_2}\geq 1}
(m^2)^{Ld/2 - \Sigma n_i+N_2} f(n_1,\ldots,n_{N_1+N_2};d)
\prod_{a=N_1+1}^{N_1+N_2}
D_a^{n_a-1}\,.
\label{integral1}
\eea
In particular, the `on-shell' value at $p_i\cdot p_k=\mu_{ik}m^2$
of the integral is represented
by the first term $f(n_1,..,n_{N_1},1,..,1;d)$ in the expansion in
$D_{ik}=p_i\cdot p_k-\mu_{ik}m^2$, while the other terms in the
expansion play an  auxiliary role when proving the equivalence.

Acting by
$(\partial/\partial p_i)\cdot p_k$,  $(i=1,\dots,L; k=1,\dots,L+R)$
on the integrand in (\ref{integral1}) we obtain $N_1$ IBP relations
exactly in the `vacuum' form (\ref{rr}).
To control the evolution of the $N_2$ `external' indices we need
additional $N_2$ relations. We obtain them by acting by
$p_k \cdot (\partial/\partial p_i)$, $(i,k=L+1,\dots,L+R)$ on
(\ref{integral1}).
These new relations look like (\ref{rr})  with the only exception that
they have no term on the left-hand side proportional to the space-time
dimension $d$.
To transform these additional relations into a manifestly equivalent form
we rescale the initial integrals by a power of the  determinant
of external kinematics invariants normalized on its `on-shell' value
$D\equiv\det(p_i\cdot p_k)/\det(\mu_{ik}m^2)\,, \quad L<i,k\leq L+R$:
\be
 F(n_1,\ldots,n_N;d;\ldots)
=D^{(R+1-d)/2}
\tilde{F}(n_1,\ldots,n_N;d; \ldots) \,,
\label{resc}
\ee
with $\tilde{F}$ expanded in $\tilde{f}$
by the same equation (\ref{integral1}) as in the case of $F$.
For pure  `on-shell' values, the prefactor in the right-hand side of 
(\ref{resc})
is equal to one. For higher terms of the expansion in the external invariants,
this prefactor is expanded and 
provides a linear substitution (easily invertible) of coefficients
$f$ throught $\tilde{f}$.
As a result, the coefficients
$\tilde{f}(n_1,\ldots,n_{N_1+N_2};d)$
exactly satisfy the `vacuum'  relations (\ref{rr}) with $N=N_1+N_2$.

Of course, boundary conditions for the recurrence procedures can be
different. In particular, when solving the recurrence
relations for pure the `on-shell' $L$-loop $(R+1)$-point integrals we can use
the combinatorial results for the corresponding $(L+R)$-loop vacuum or
$(L+R-1)$-loop propagator recurrence relations, but with the additional
condition that integrals with non-positive values of the `external'
indices should be put to zero.

\section{Recurrence procedure for massless 2-loop vertex diagrams}

Let us apply the equivalence property described in the previous section
to the evaluation of the massless two-loop vertex diagrams  shown
\begin {figure} [ht]
\begin{picture}(200,140)(-30,-80)
\Line(0,0)(35,20)
\Line(35,-20)(0,0)
\Line(35,20)(35,-20)
\Line(35,20)(70,40)
\Line(70,-40)(35,-20)
\Line(70,-40)(70,40)
\ArrowLine(-25,0)(0,0)
\ArrowLine(70,40)(88,50.28570)
\ArrowLine(70,-40)(88,-50.28570)
\Vertex(0,0){1.5}
\Vertex(35,20){1.5}
\Vertex(35,-20){1.5}
\Vertex(70,40){1.5}
\Vertex(70,-40){1.5}
\Text(-10,10)[]{\small $q$}
\Text(79,55)[]{\small $p_1$}
\Text(79,-55)[]{\small $p_2$}
\Text(17.5,20)[]{\small $1$}
\Text(17.5,-20)[]{\small $2$}
\Text(52.5,40)[]{\small $3$}
\Text(52.5,-40)[]{\small $4$}
\Text(77,0)[]{\small $5$}
\Text(42,0)[]{\small $6$}
\Text(35,-65)[]{\small (a)}
\Line(140,0)(175,20)
\Line(175,-20)(140,0)
\Line(175,20)(210,40)
\Line(210,-40)(175,-20)
\Line(175,20)(210,-40)
\Line(175,-20)(210,40)
\ArrowLine(115,0)(140,0)
\ArrowLine(210,40)(228,50.28570)
\ArrowLine(210,-40)(228,-50.28570)
\Vertex(140,0){1.5}
\Vertex(175,20){1.5}
\Vertex(175,-20){1.5}
\Vertex(210,40){1.5}
\Vertex(210,-40){1.5}
\Text(130,10)[]{\small $q$}
\Text(219,55)[]{\small $p_1$}
\Text(219,-55)[]{\small $p_2$}
\Text(175,-65)[]{\small (b)}
\Line(280,0)(315,20)
\Line(315,-20)(280,0)
\Line(315,20)(350,0)
\Line(315,20)(350,40)
\Line(350,-40)(315,-20)
\Line(350,-40)(350,40)
\ArrowLine(255,0)(280,0)
\ArrowLine(350,40)(368,50.28570)
\ArrowLine(350,-40)(368,-50.28570)
\Vertex(280,0){1.5}
\Vertex(315,20){1.5}
\Vertex(350,0){1.5}
\Vertex(350,40){1.5}
\Vertex(350,-40){1.5}
\Text(270,10)[]{\small $q$}
\Text(359,55)[]{\small $p_1$}
\Text(359,-55)[]{\small $p_2$}
\Text(315,-65)[]{\small (c)}
\end{picture}
\caption{Three types of two-loop vertex diagrams with
general numerators
and integer powers of propagators: (a) planar, (b) non-planar and
(c) non-Abelian.}
\label{2lver}
\end{figure}
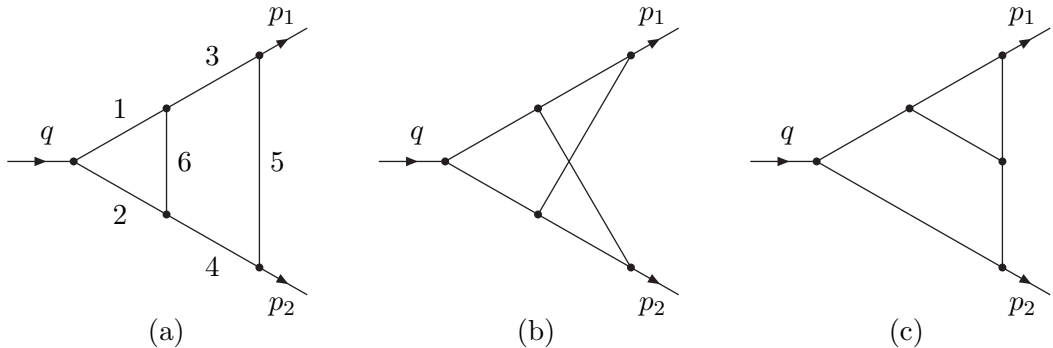
in Fig.~\ref{2lver}a,b,c with two legs on the light cone, $p_1^2=p_2^2=0$,
general polynomials in the numerator and arbitrary integer indices of
the propagators. The diagrams depend homogeneously on
$q^2=(p_1+p_2)^2 =-Q^2$.
 For example, a general Feynman integral
for the planar graph Fig.~\ref{2lver}a  can be written as
\bea
 F_p(Q,\ep)  &=&
 \int  \int \frac{\dd^dk \dd^dl}{(l^2-2 p_1\cdot  l)^{n_1}
(l^2+2 p_2\cdot l)^{n_2}}
 \nn \\ & & 
 \times \frac{(k\cdot l)^{n_7}}{
(k^2-2 p_1\cdot k)^{n_3} (k^2+2 p_2\cdot k)^{n_4}
( k^2)^{n_5} ((k-l)^2)^{n_6}} \, .
\label{F1}
\eea
The standard $i 0$ prescription is implied, i.e. $ k^2= k^2+i 0$.
We have chosen the irreducible numerator to be the scalar product of the
two loop momenta, $k\cdot l$.

 The planar and non-planar integrals with
the powers of propagators equal to one and some numerators
of low degree were evaluated in \cite{Gons} in expansion in $\ep$.
The planar integral with the powers of propagators equal to one
was evaluated  by IBP in gamma functions for general $\ep$
in \cite{KL}.
Using IBP, an algorithm for the evaluation of general planar integrals
without numerators has been developed
when expanding two-loop vertex diagrams in the Sudakov
limit \cite{Sm1}. This solution of the recurrence relations
was reproduced in \cite{DOT}. For planar diagrams with arbitrary
numerators, a complete solution was presented in \cite{Sm3}.
The non-Abelian diagrams Fig.~\ref{2lver}c are non more complicated
than the planar ones. The IBP relations for them are simple, and
any diagram can be expressed in gamma functions for general
values of $d$. Thus the only non-trivial case is in fact the
non-planar diagram of Fig.~\ref{2lver}b.

\begin {figure} [ht]
\begin{picture}(200,90)(-30,-60)
\Line(10,0)(-5,0)
\CArc(30,0)(20,90,270)
\Line(30,20)(30,-20)
\Line(30,20)(50,20)
\Line(30,-20)(50,-20)
\CArc(50,0)(20,-90,90)
\Line(50,20)(50,-20)
\Line(70,0)(85,0)
\Vertex(10,0){1.5}
\Vertex(30,20){1.5}
\Vertex(30,-20){1.5}
\Vertex(50,20){1.5}
\Vertex(50,-20){1.5}
\Vertex(70,0){1.5}
\Text(42,-40)[]{\small (a)}
\Line(135,0)(150,0)
\CArc(170,0)(20,90,270)
\Line(170,20)(190,20)
\Line(170,-20)(190,-20)
\CArc(190,0)(20,-90,90)
\Line(170,20)(190,-20)
\Line(190,20)(170,-20)
\Line(210,0)(225,0)
\Vertex(150,0){1.5}
\Vertex(170,20){1.5}
\Vertex(170,-20){1.5}
\Vertex(190,20){1.5}
\Vertex(190,-20){1.5}
\Vertex(210,0){1.5}
\Text(182,-40)[]{\small (b)}
\Line(290,0)(275,0)
\CArc(310,0)(20,90,270)
\Line(310,20)(330,20)
\Line(310,-20)(330,-20)
\CArc(330,0)(20,-90,90)
\Line(320,-20)(320,0)
\Line(310,20)(320,0)
\Line(330,20)(320,0)
\Line(350,0)(365,0)
\Vertex(290,0){1.5}
\Vertex(310,20){1.5}
\Vertex(320,-20){1.5}
\Vertex(330,20){1.5}
\Vertex(320,0){1.5}
\Vertex(350,0){1.5}
\Text(322,-40)[]{\small (c)}
\end{picture}
\caption{Three types of three-loop propagator diagrams with
general numerators
and integer powers of propagators: (a) planar, (b) non-planar and
(c) Mercedes.}
\label{3lprop}
\end{figure}
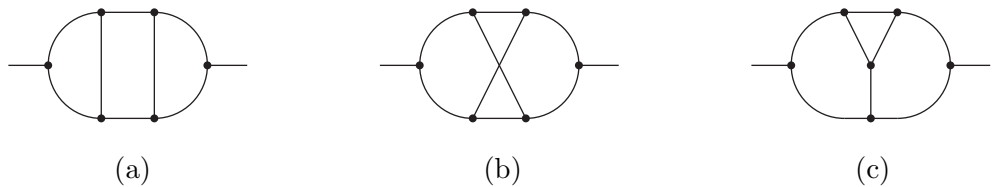

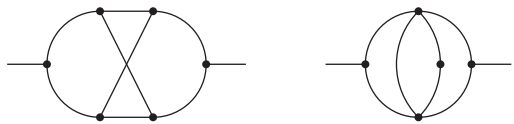
\begin {figure} [ht]
\begin{picture}(200,90)(-30,-35)
\Line(65,0)(80,0)
\CArc(100,0)(20,90,270)
\Line(100,20)(120,20)
\Line(100,-20)(120,-20)
\CArc(120,0)(20,-90,90)
\Line(100,20)(120,-20)
\Line(120,20)(100,-20)
\Line(140,0)(155,0)
\Vertex(80,0){1.5}
\Vertex(100,20){1.5}
\Vertex(100,-20){1.5}
\Vertex(120,20){1.5}
\Vertex(120,-20){1.5}
\Vertex(140,0){1.5}
\Line(200,0)(185,0)
\CArc(220,0)(20,0,360)
\CArc(200,0)(28.2843,-45,45)
\CArc(240,0)(28.2843,135,225)
\Line(240,0)(255,0)
\Vertex(200,0){1.5}
\Vertex(220,20){1.5}
\Vertex(220,-20){1.5}
\Vertex(240,0){1.5}
\Vertex(228.2843,0){1.5}
\end{picture}
\caption{Master diagrams for 3-loop propagators.}
\label{3lpropMaster}
\end{figure}

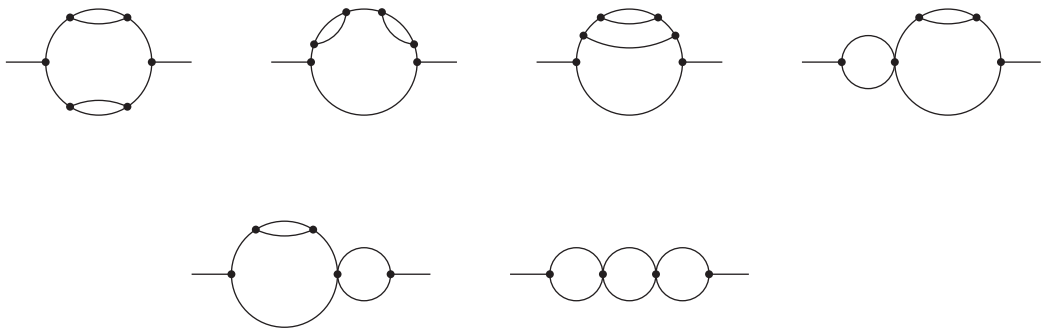
\begin {figure} [ht]
\begin{picture}(200,155)(-30,-35)
\Line(-15,80)(0,80)
\CArc(20,80)(20,0,360)
\CArc(20,120)(25.5645,-115,-65)
\CArc(20,40)(25.5645,65,115)
\Line(40,80)(55,80)
\Vertex(0,80){1.5}
\Vertex(30.804,96.8307){1.5}
\Vertex(30.804,63.1693){1.5}
\Vertex(9.19599,96.8307){1.5}
\Vertex(9.19599,63.1693){1.5}
\Vertex(40,80){1.5}
\Line(85,80)(100,80)
\CArc(120,80)(20,0,360)
\CArc(100,100)(13.3328,-85,-5)
\CArc(140,100)(13.3328,185,265)
\Line(140,80)(155,80)
\Vertex(100,80){1.5}
\Vertex(101.16203,86.7179){1.5}
\Vertex(113.2821,98.838){1.5}
\Vertex(126.718,98.838){1.5}
\Vertex(138.838,86.7179){1.5}
\Vertex(140,80){1.5}
\Line(185,80)(200,80)
\CArc(220,80)(20,0,360)
\CArc(220,120)(25.5645,-115,-65)
\CArc(220,120)(34.641,-120,-60)
\Line(240,80)(255,80)
\Vertex(200,80){1.5}
\Vertex(230.804,96.8307){1.5}
\Vertex(202.679,90){1.5}
\Vertex(209.19599,96.8307){1.5}
\Vertex(237.321,90){1.5}
\Vertex(240,80){1.5}
\Line(285,80)(300,80)
\CArc(310,80)(10,0,360)
\CArc(340,80)(20,0,360)
\CArc(340,120)(25.5645,-115,-65)
\Line(360,80)(375,80)
\Vertex(320,80){1.5}
\Vertex(300,80){1.5}
\Vertex(350.804,96.8307){1.5}
\Vertex(329.19599,96.8307){1.5}
\Vertex(360,80){1.5}
\Line(70,0)(55,0)
\Line(130,0)(145,0)
\CArc(90,0)(20,0,360)
\CArc(90,40)(25.5645,-115,-65)
\CArc(120,0)(10,0,360)
\Vertex(100.804,16.8307){1.5}
\Vertex(79.19599,16.8307){1.5}
\Vertex(70,0){1.5}
\Vertex(110,0){1.5}
\Vertex(130,0){1.5}
\Line(190,0)(175,0)
\Line(250,0)(265,0)
\CArc(200,0)(10,0,360)
\CArc(220,0)(10,0,360)
\CArc(240,0)(10,0,360)
\Vertex(190,0){1.5}
\Vertex(210,0){1.5}
\Vertex(230,0){1.5}
\Vertex(250,0){1.5}
\end{picture}
\caption{
Recursively one-loop  diagrams for 3-loop propagators. Integer indices of lines
and  numerators are arbitrary.}
\label{3lpropRec1l}
\end{figure}

 Following the results of Sect.~2 we reduce the evaluation of
the diagrams of  Fig.~\ref{2lver} to the evaluation of
three-loop massless propagator diagrams of
 Fig.~\ref{3lprop}a,b,c, with
arbitrary numerators and integer powers of propagators.
As is well known the latter can be reduced, using the algorithm
of \cite{IBP}, to a linear combination of
master integrals Fig.~\ref{3lpropMaster} and recursively one-loop integrals
 Fig.~\ref{3lpropRec1l}

Using the statement justified in the previous section we
conclude that to solve IBP relations for Fig.~\ref{2lver}
we can use recurrence relations for Fig.~\ref{3lprop}
presented in \cite{IBP} in an explicitly  solved form.
The massless external legs should be identified with
a pair of lines of the propagator diagrams which are connected to
an (arbitrarily fixed) external vertex. The  additional
condition formulated above means that, when using the recurrence
procedure, one should put to zero integrals with non-positive values of
these `former external' indices of the lines.

To make a correspondence between the two given families of
the diagrams let us choose, for definiteness,
the right vertex in Fig.~\ref{3lprop}a,b,c and lines that are
incident to this vertex so that
the vertex diagrams of  Fig.~\ref{2lver} will be obtained from
the propagator diagrams of Fig.~\ref{3lprop} by cutting
the two right internal lines and, vice versa, the propagator
diagrams will be obtained from the vertex diagrams by adding
a new external vertex and a pair of lines which connect them
with the two right external vertices in Fig.~\ref{2lver}.

Thus, to evaluate a vertex diagram from Fig.~\ref{2lver},
with a given set of indices $n_i$,
we can use the following prescriptions:
\begin{itemize}
\item
using the algorithm presented in \cite{IBP} for
the corresponding 3-loop propagator analog
express it as a linear combination of the master integrals of
 Fig.~\ref{3lpropMaster} and recursively one-loop diagrams
Fig.~\ref{3lpropRec1l};
\item
omit the contribution of the recursively one-loop diagrams in the
first raw of Fig.~\ref{3lpropRec1l};
\item
substitute the values of the recursively one-loop diagrams from the second
raw in Fig.~\ref{3lpropRec1l} by
the corresponding values of the recursively one-loop  vertex-type integrals
shown in Fig.~\ref{2lverRec1l} if the  indices of the right two lines are
equal  to one. If at least one of the right indices is greater than one, then
evaluate the propagator diagram and express the result (written in gamma
functions) through the corresponding propagator diagram with both indices
equal to one. Finally, perform the above substitution in this result;
\item
substitute the values of the  master integrals in Fig.~\ref{3lpropMaster}
by the corresponding values of the master integrals shown in
 Fig.~\ref{2lverMaster}.
\end{itemize}

\begin {figure} [ht]
\begin{picture}(200,90)(-30,-50)
\Line(80,0)(115,20)
\Line(115,-20)(80,0)
\Line(115,20)(150,40)
\Line(150,-40)(115,-20)
\Line(115,20)(150,-40)
\Line(115,-20)(150,40)
\Line(65,0)(80,0)
\Line(150,40)(168,50.28570)
\Line(150,-40)(168,-50.28570)
\Vertex(80,0){1.5}
\Vertex(115,20){1.5}
\Vertex(115,-20){1.5}
\Vertex(150,40){1.5}
\Vertex(150,-40){1.5}
\Line(260,0)(295,20)
\Line(260,0)(295,-20)
\Line(260,0)(305,25.7143)
\Line(260,0)(305,-25.7143)
\CArc(275,0)(28.2843,-45,45)
\CArc(315,0)(28.2843,135,225)
\Line(245,0)(260,0)
\Vertex(260,0){1.5}
\Vertex(295,20){1.5}
\Vertex(295,-20){1.5}
\Vertex(303.284,0){1.5}
\end{picture}
\caption{
Master diagrams for 2-loop vertices. All the lines have indices equal
to one and the numerator is equal to one.}
\label{2lverMaster}
\end{figure}
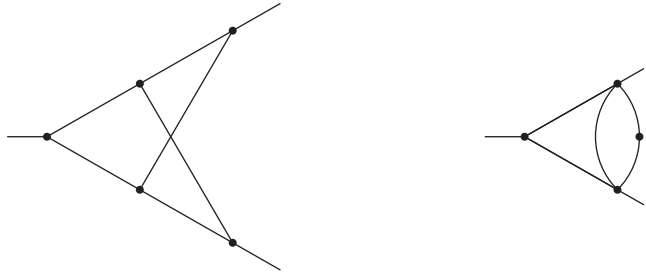
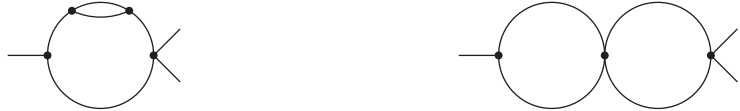
\begin {figure} [ht]
\begin{picture}(200,80)(-30,-40)
\Line(90,0)(75,0)
\Line(130,0)(140,10)
\Line(130,0)(140,-10)
\CArc(110,0)(20,0,360)
\CArc(110,40)(25.5645,-115,-65)
\Vertex(120.804,16.8307){1.5}
\Vertex(99.19599,16.8307){1.5}
\Vertex(90,0){1.5}
\Vertex(130,0){1.5}
\Line(245,0)(260,0)
\Vertex(260,0){1.5}
\Vertex(300,0){1.5}
\Vertex(340,0){1.5}
\CArc(280,0)(20,0,360)
\CArc(320,0)(20,0,360)
\Line(340,0)(350,10)
\Line(340,0)(350,-10)
\end{picture}
\caption{Recursively one-loop diagrams for 2-loop vertices.
Integer indices of lines and  numerators are arbitrary.}
\label{2lverRec1l}
\end{figure}

The correspondence between the two sets of the integrals is obvious:
the diagrams from  Fig.~\ref{2lverMaster} and~\ref{2lverRec1l}
are obtained respectively from  Fig.~\ref{3lpropMaster} and~\ref{3lpropRec1l}
by cutting a pair of the right lines.
Note however that the second master vertex integral in
 Fig.~\ref{2lverMaster} is in fact recursively one-loop and easily evaluated
in gamma functions for general $\ep\equiv (4-d)/2$ so that the only `true' 
master
vertex integral is the non-planar diagram in  Fig.~\ref{2lverMaster} which
was evaluated in \cite{Gons} in expansion in $\ep$ up to the finite
part:
\bea
\int  \int \frac{\dd^dk \dd^dl}{((k+l)^2-2 p_1 (k+l))
((k+l)^2+2 p_2 (k+l))
(k^2-2 p_1 k) (l^2+2 p_2 l) k^2 l^2} \nn \\
= \left(
\frac{1}{\ep^4} - \frac{\pi^2}{\ep^2}  -
  \frac{83 \zeta(3)}{3 \ep} - \frac{59 \pi^4}{120}
\right) \frac{\left(i\pi^{d/2} e^{-\gm_E \ep}\right)^2}{(Q^2)^{2+2\ep}} \, ,
\label{NP}
\eea
where $\gm_E$ is the Euler constant.

\vspace{2mm}

{\em Acknowledgments.}
This work was supported by the Volkswagen Foundation, contract
No.~I/73611 and by the Russian Foundation for Basic
Research, project 98--02--16981. P.A.B.  also acknowledges support of
INTAS, grant YSF-98-173.
We are grateful to K.G.~Chetyrkin for stimulating discussions and helpful
comments on draft versions of the paper. We are thankful to Z.~Kunszt for
stimulating discussions.

\end{document}